# Coherent mode-combined ultra-narrow-linewidth single-mode micro-disk laser

Jintian Lin,[1,7,#] Saeed Farajollahi,[2,#] Zhiwei Fang,[3,#] Ni Yao,[4,5] Renhong Gao,[1,7] Jianglin Guan,[3,6] Li Deng,[3,6] Tao Lu,[2,†] Min Wang,[2,6] Haisu Zhang,[3,6] Wei Fang,[5,‡] Lingling Qiao,[1,7] and Ya Cheng[1,3,6,7,8,9,*]

[1]*State Key Laboratory of High Field Laser Physics and CAS Center for Excellence in Ultra-Intense Laser Science, Shanghai Institute of Optics and Fine Mechanics (SIOM), Chinese Academy of Sciences (CAS), Shanghai 201800, China.*

[2]*Department of Electrical and Computer Engineering, University of Victoria, Victoria, British Columbia V8P 5C2, Canada.*

[3]*XXL—The Extreme Optoelectromechanics Laboratory, School of Physics and Electronic Science, East China Normal University, Shanghai 200241, China.*

[4]*Research Center for Intelligent Sensing, Zhejiang Lab, Hangzhou 311100, China.*

[5]*The Interdisciplinary Center for Quantum Information, State Key Laboratory of Modern Optical Instrumentation, College of Optical Science and Engineering, Zhejiang University, Hangzhou 310027, China.*

[6]*State Key Laboratory of Precision Spectroscopy, East China Normal University, Shanghai 200062, China.*

[7]*Center of Materials Science and Optoelectronics Engineering, University of Chinese Academy of Sciences, Beijing 100049, China.*

[8]*Collaborative Innovation Center of Extreme Optics, Shanxi University, Taiyuan 030006, China.*

[9]*Collaborative Innovation Center of Light Manipulations and Applications, Shandong Normal University, Jinan 250358, China.*

[10]*Shanghai Research Center for Quantum Sciences, Shanghai 201315, China.*

[#]*These authors contributed equally to the work.*

[†]Electronic address: taolu@ece.uvic.ca

[‡]Electronic address: wfang08@zju.edu.cn

[*]Electronic address: ya.cheng@siom.ac.cn

Date: July 19, 2021

**Integrated single-mode microlasers with ultra-narrow linewidths play a game-changing role in a broad spectrum of applications ranging from coherent communication and LIDAR to metrology and sensing. Generation of such light sources in a controllable and cost-effective manner remains an outstanding challenge due to the difficulties in the realization of ultra-high Q active micro-resonators with suppressed mode numbers. Here, we report a microlaser generated in an ultra-high Q Erbium doped lithium niobate (LN) micro-disk. Through the formation of coherently combined polygon modes at both pump and laser wavelengths, the microlaser exhibits single mode operation with an ultra-narrow-linewidth of 98 Hz. In combination with the superior electro-optic and nonlinear optical properties of LN crystal, the mass-producible on-chip single-mode microlaser will provide an essential building block for the photonic integrated circuits demanding high precision frequency control and reconfigurability.**

With a broad transparency window as well as high piezoelectric, acousto-optic, second order nonlinear and electro-optic coefficients, crystalline lithium niobate (LN) has been considered as the "silicon in photonics" [1-7]. Thanks to the recent breakthroughs in the nanofabrication technology [4], various thin film LN integrated photonic devices such as high-performance electro-optic modulators [8,9], broad-band optical frequency combs [10-12], and high-efficient frequency convertors [13-15], etc., have been demonstrated. However, similar to silicon, LN itself cannot provide optical gain that hinders it from generating laser directly from LN chip. Doping of Erbium ion ($Er^{3+}$) into the LN substrate makes the on-chip microlaser at C-band possible [16-21]. Nevertheless, realization of single-mode microlasers with narrow linewidths in thin film LN is still a challenge due to the inherent multi-mode nature in high-Q whispering gallery mode (WGM) optical microcavities.

According to the Schawlow-Townes theory, the laser fundamental linewidth is quadratic inverse proportional to the undoped cavity Q factor [27-29]. Consequently, increasing the Q factor will quadratically reduce the linewidth of the microlaser at a given output power, and benefit the photonic applications. The highest Q factors demonstrated to date are those of WGM microcavities where the confinement of light is achieved by the continuous total internal reflection at the smooth resonator interface [30]. However, the dense WGMs within the gain spectrum would easily lead to multi-mode lasing in a micro-disk. To reduce the number of modes, attempts have been made to reduce the microcavity size[22], which inevitably increase the radiation loss and decrease the mode volume leading to reduction of both the Q and gain in the microcavity. Consequently, the pump threshold power must increase while the laser power still remains low.

In this work, taking the advantage of the coherently combined polygon modes in high-Q micro-disks formed by weak perturbation from a tapered fiber [31], we propose a novel approach for single-mode narrow-linewidth microlasers in single WGM micro-disks. To verify this approach, an $Er^{3+}$ doped LN micro-disk with a diameter of ~29.8 μm is fabricated, and a fiber taper with a diameter of 2 μm is used for coupling light beams into and out of the micro-disk. The polygon modes are formed from the coherent combination of normal WGMs through the weak perturbation of the tapered optical fiber with minimum impairments on loaded Q factors of the polygon modes. Because the polygon modes are spatially distributed far from the periphery of the disk, and in turn less affected by the scattering loss at the air/LN boundary, the Q factors of which can reliably reach above $10^7$. Single-mode lasing is observed from a polygon mode at 1546 nm wavelength when pumping the micro-disk at 968 nm wavelength. Despite the large disk diameter of almost 30 μm, the single-mode lasing is observed at a threshold pump power as low as 25 μW with a linewidth as narrow as 98.0 Hz. The micro-disk laser operates stably at the room temperature.

A scanning electron microscope (SEM) image of the fabricated microcavity is shown in the inset of Fig. 1, which consists of an $Er^{3+}$ doped Z-cut LN micro-disk with a diameter of 29.8 μm and a thickness of 700 nm. The micro-disk is supported by a thin silica pedestal (with a diameter of 7.9 μm) supported by the LN substrate. The doping concentration of $Er^{3+}$ is chosen as 1 mol.%. The fabrication process of the micro-disk can be found in the **Methods** section. The experimental setup is schematically illustrated in Fig. 1. A narrow linewidth tunable diode laser (Model: TLB-6719, New Focus Inc.) was used as the pump light source, and the pump laser wavelength was tuned to be resonant with a specific transverse electric (TE) polarized cavity mode around 968 nm. The laser was coupled to the micro-disk via a tapered fiber with waist size of 2 μm. The position between the tapered fiber and the micro-disk was adjusted by a three-dimensional (3D) piezo stage with resolution of 20 nm. An optical imaging system consisting of a microscope objective lens with numerical aperture (NA) of 0.42, filters, and an infrared (IR) charge coupled device (CCD) was used to monitor the coupling system from top and capture the intensity profile of the modes. The specifications of the filters used to capture the intensity profiles of pump light, laser signal, and up-conversion fluorescence, can be found in the **Measurement** section. The generated emission from the micro-disk was coupled out by the same tapered fiber and sent into either an optical spectrum analyzer (OSA, detection range: 600 nm to 1700 nm) or a fiber spectrometer (detection range: 300 nm to 1000 nm) for spectral analysis. Meanwhile, the polarization states of the pump and induced light were checked via a pre-calibrated wire grating polarizer (WGP) by collecting the signals scattered from the edge of the micro-disk.

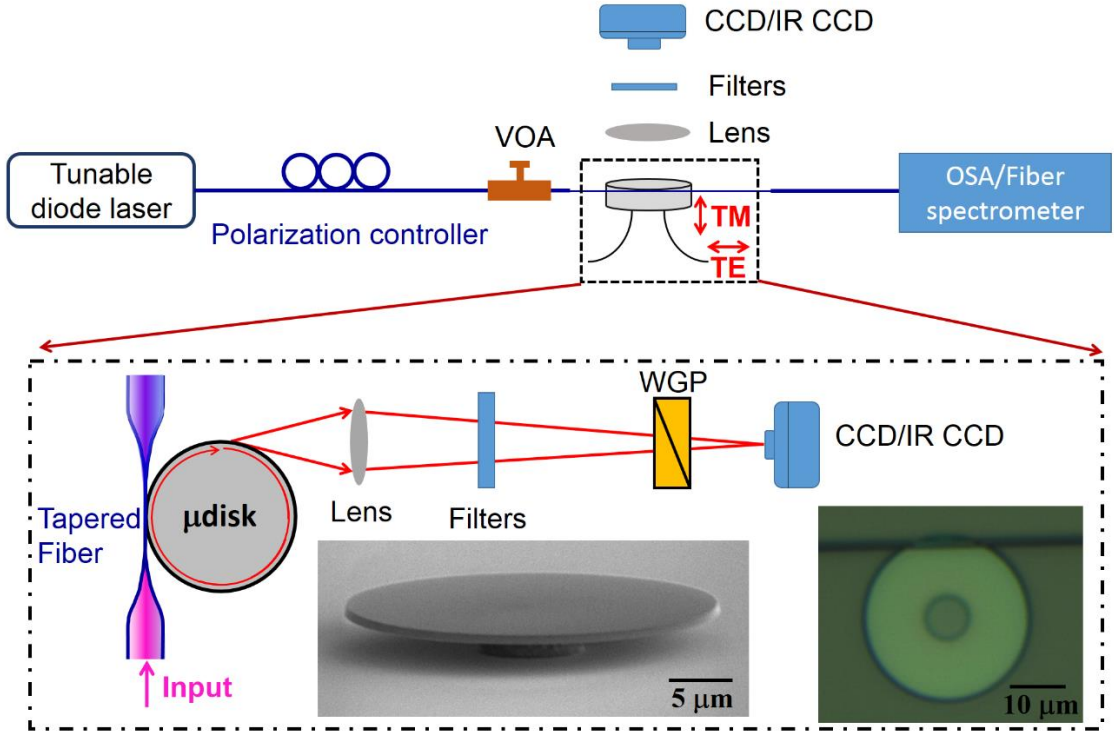

FIG. 1: **The experimental setup for lasing and polarization measurement.** Variable optical attenuator (VOA) was used tuned the input power into the micro-disk. TM and TE are shorted for transverse magnetic and transverse electric, respectively. **Inset:** Scanning electron microscope (SEM) image of the micro-disk and optical micrograph of the tapered fiber coupled with the micro-disk.

When the relative position between the tapered fiber and the center of the micro-disk was adjusted to 13.37 μm (inset in Fig. 1) and the pump laser wavelength was tuned to 968 nm, single mode laser peaked at 1546 nm wavelength was observed, as shown in Fig. 2(a). The IR CCD captured the top view of the lasing mode, as shown in Fig. 2(b), which indicates a quadrangular pattern in the micro-disk. This pattern coincides with the polygon modes formed at 968 nm and ~550 nm wavelengths as shown in the insets in Fig. 2(b). Obviously, the pump laser at 968 nm was resonant with the cavity mode. Meanwhile, the green light around 550 nm was the up-conversion fluorescence of $Er^{3+}$ excited along the trajectory of the pump light, as illustrated in the spectrum in Fig. 2(c). Remarkably, the square mode formed at the pump wavelength is clean being rather isolated from the conventional WGM modes circulating around the circumference of the disk. The effect efficiently suppresses the high-order radial modes and gives rise to the relatively large free spectral range (FSR) of 11.5 nm which facilitates single-mode lasing with the limited gain spectrum of $Er^{3+}$ ion. The output power of laser signal as a function of pump power is plotted in Fig. 2(d), indicating a lasing threshold of merely ~25 µW. In addition, the wavelength of the lasing signal is red shifted with the increasing pump power because of the significant photorefractive and thermo-optic effect [17,32,33], as shown in Fig. 2(e). The wavelength tuning rate is 0.59 nm/mW (i.e., 74 MHz/µW). And the maximum output power of the microlaser was measured as 2 µW.

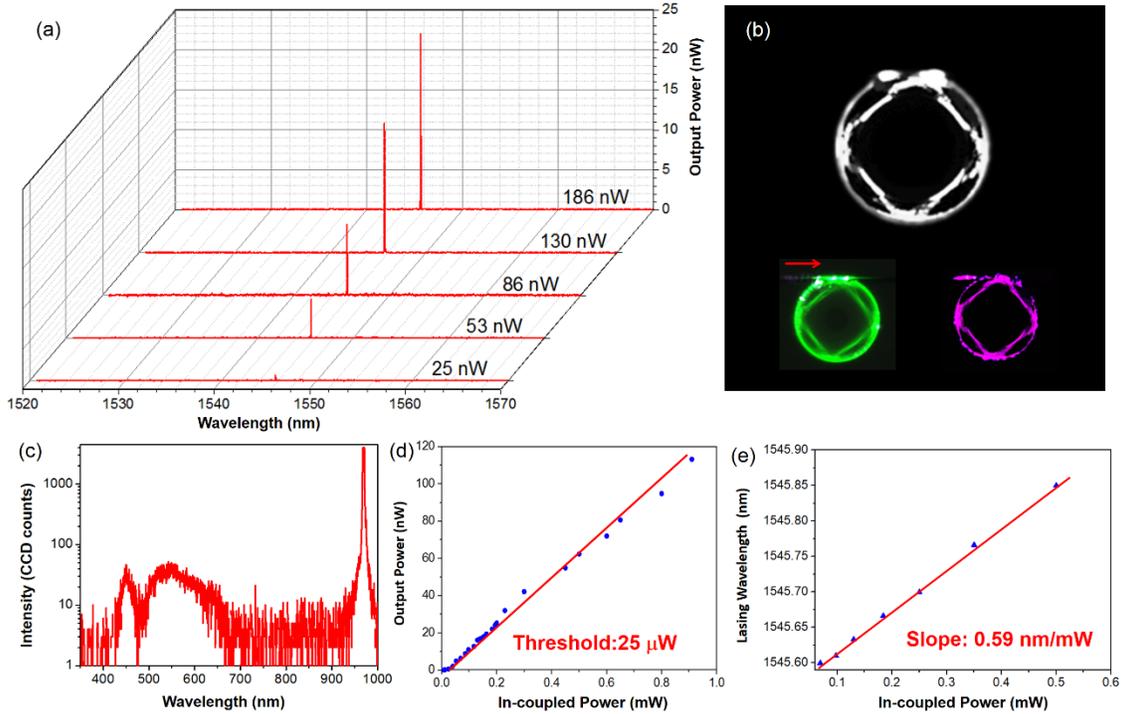

FIG. 2: **Spectra of lasing**. **a,** Spectra of the output power of the microlaser varied with the pump power. **b,** The optical micrograph of the square lasing modes at 1546 nm wavelength. Inset: the optical micrographs of the square modes of the up-conversion fluorescence around 550 nm wavelength (Left) and the pump light (Right). **c,** Spectrum of the up-conversion fluorescence and the pump light. **d,** Power dependent output power of single-mode lasing. **e,** The pump-power dependent wavelength of the lasing signal.

To characterize the Q factors of the relevant modes in undoped microcavities, a micro-disk with the same geometry as the doped microcavity was fabricated. More interestingly, the square modes show higher Q-factors in comparison to the WGMs distributed around the circumference. Figures 3(a) and (d) show the transmission spectra around 968 nm and 1546 nm when two tunable diode lasers (Model: TLB-6719 & 6728, New Focus Inc.) with an output power of 5 µW were coupled into the undoped micro-disk via the tapered fiber at the same position, respectively. The Q factors of the pump laser mode at 968 nm (Fig. 3(b)) and the lasing mode at 1546 nm (Fig. 3(e)) were measured to be above $10^7$, about one order of magnitude higher than conventional WGMs in the vicinity (Figs. 3(c) and (f)). This can be understood as the square modes are distributed further away from the periphery of the disk than the conventional WGMs, by which the scattering loss can be avoided.

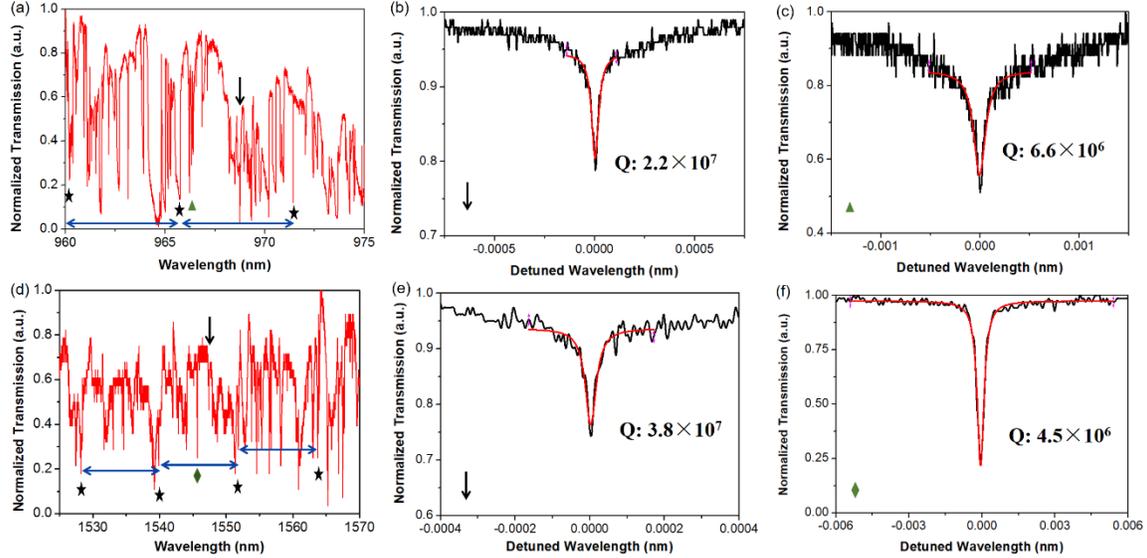

FIG. 3: **Transmission spectra of the tapered fiber coupled with the micro-disk**. **a**. Transmission spectrum around pump wavelength, where each blue line segment with 2 arrows indicates an FSR. **b**. The Q factor of the pump mode. **c**. The measured highest Q factor of the conventional WGM (indicated with a yellow star at Fig. 3(a)) within one FSR. **d**. Spectrum around lasing wavelength. **e**. The Q factor of the lasing mode. **f**. The measured highest Q factor of the conventional WGM (indicated with green rhombus in Fig. 3(d)) within one FSR.

The linewidth was measured by heterodyning two independently pumped microlasers at slightly different lasing frequencies (with a difference of ~MHz) [34]. The two micro-disks possess the same diameter, and the lasing wavelengths are shifted with 1 MHz when the in-coupled pump power of one micro-disk is ~13.5 µW higher than that of the other micro-disk. The experimental setup is presented in Fig. 4(a), where a polarization controller is used to ensure the two laser signals share the same polarization state. The two laser signals are combined with a 3 dB coupler and sent into a balanced photodetector (PD). A Lorentzian-shaped power spectrum of the beat note is observed in the real time spectrum analyzer (Tektronix RSA5126B) with a resolution of ~1 Hz. Figure 4(b) shows the radio frequency spectrum when the output power of both microlasers are ~2 µW. The full width at half-maximum (FWHM) of the detected signal is obtained as 196.0 Hz, through a least square fitting to a Lorentzian function. Therefore, at 2 µW output power, each laser signal from the micro-disk possesses a fundamental linewidth of 98.0 Hz. Moreover, the power spectral density of the frequency noise is also recorded by the spectrum analyzer, as shown in Fig. 4(c). At high frequency range (> 0.5 MHz), the frequency noise is almost flat around 194.2 Hz$^2$/Hz, which indicates a fundamental linewidth of ~97.1 Hz for each microlaser and is consistent with the result in Fig. 4(b). The linewidth of the microlaser as a function of the output power was further characterized. The dependence of the linewidth on output power agrees well with the Schawlow-Townes formula [27-29], i.e., the linewidth is inversely proportional to the laser output power.

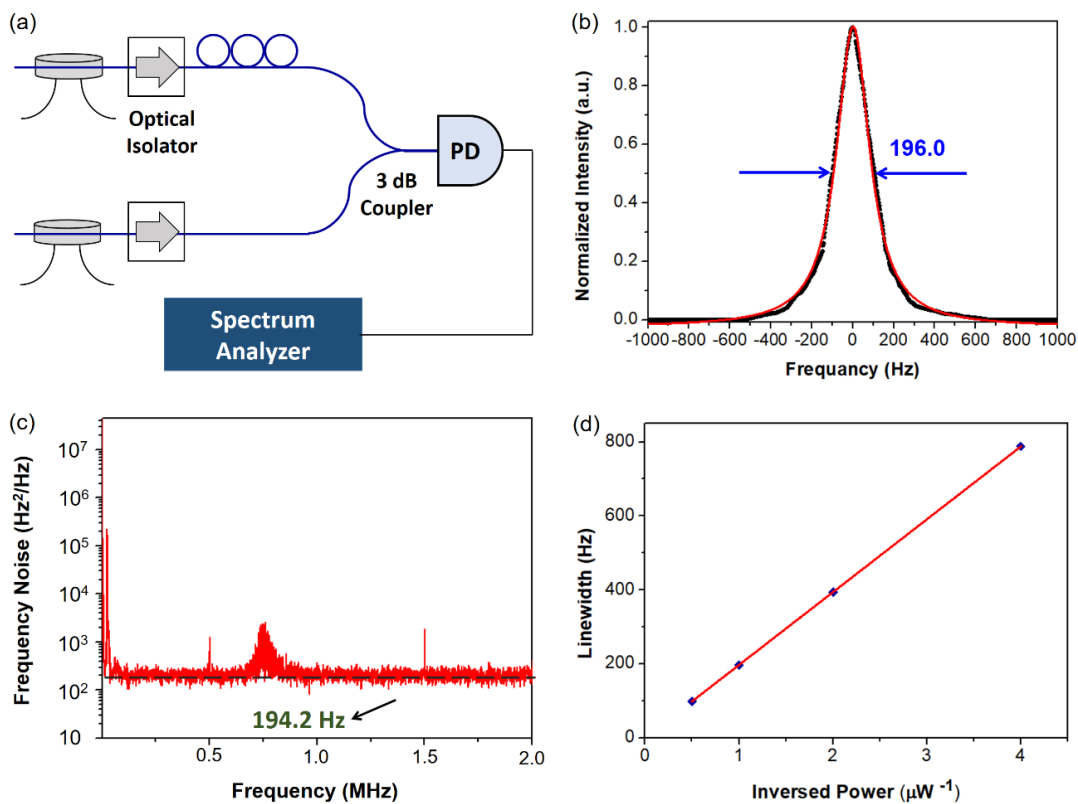

FIG. 4: **Linewidth measurement**. **a**. The experimental setup. **b**. The Lorentz fitting (red line) of the detected beating signal (black dots) for output power of 2 μW. **c**. The measured frequency noise. **d**. The linewidth as a function of the output power of the microlaser.

The underlying physics behind the polygon modes can be explained by performing a 2D finite element simulation using COMSOL. The extraordinary and ordinary refractive index of LN used in the simulation are $n_e$ = 2.1622 and $n_o$ = 2.2393 for the pump wavelength and $n_e$ = 2.1361 and $n_o$ = 2.2113 for the laser wavelength [35]. To provide sufficient perturbation for polygon modes, a tapered fiber is placed on top of the disk and we use the effective index method to compute the effective refractive index of the fiber-disk overlap regime [36,37]. Figures 5(a) and (b) show the intensity profiles of two square modes at the pump and laser wavelengths of 967.36 nm and 1553.21 nm. It is worth mentioning that within one FSR around each laser and pump wavelengths, multiple number of WGMs co-exist but only single square mode will appear due to the stringent requirement for polygon mode formation. We then compute the overlap factor $\Gamma$ between different laser and pump modes defined by

$$\Gamma = \frac{\iiint I_p \cdot I_l \, dv}{\sqrt{\iiint I_p^2 \, dv} \cdot \sqrt{\iiint I_l^2 \, dv}} \tag{1}$$

Here, $I_p$ and $I_l$ are pump and laser intensities and $\iiint (\cdot) dv$ is the volume integral over full space. When the pump laser is in square mode, its overlap factor with a square mode at laser wavelength is $\Gamma = 0.75$ while its overlap factor with the fundamental WGM can be as low as $\Gamma = 0.2$. Therefore, when pumped by a square mode, a square laser mode can be excited at the much lower threshold power than WGMs due to its high $\Gamma$ value. In addition, the FSR is larger than the gain bandwidth of $Er^{3+}$ ion. Therefore, the square mode at neighboring FSRs will not be excited due to lower gain. Consequently, single mode operation can be achieved when the polygon mode is formed at pump wavelength. In contrast, when a regular WGM is formed by the pump, multiple WGMs within the gain bandwidth may lase simultaneously as $\Gamma$ between laser and pump WGMs are close in value. Finally, it is also worth mentioning that the 2D simulation will not allow to precisely evaluate the overall quality factor of the coupled cavity due to the unphysically large reflection incurred at the fiber-to-disk interfaces. The issue can be overcome with a full wave 3D mode match method which will be applied in future research [36,37]. Nevertheless, the 2D simulation results capture the essence of the polygon mode microlaser which faithfully reproduces the main features of the observed micro-disk laser mode as shown below.

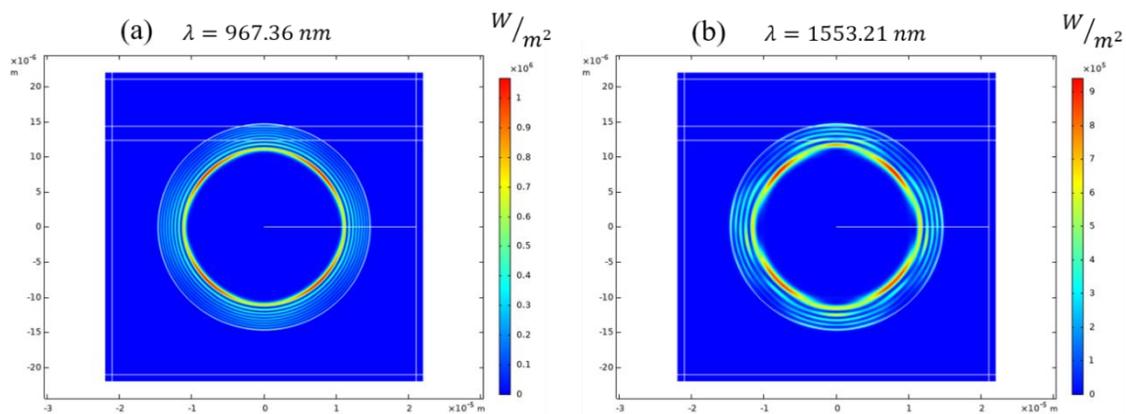

FIG. 5: **The intensity profile of square modes. a,** The intensity profile of the mode at pump wavelength of 967.36 nm. **b,** The intensity profile of the mode at laser wavelength of 1553.21 nm.

In summary, the formation of coherent polygon modes with ultra-high Q factors has allowed for the realization of single-mode narrow-linewidth microlasers in single LN micro-disks, which has significant implication for the miniaturization of optical systems in which highly coherent laser sources must be incorporated. Further exploration of the strong piezoelectric, acousto-optic, second order nonlinear and electro-optic properties of the LN will promote the performance and functionality of the single-mode micro-disk laser in a straightforward manner without the necessity of heterogeneous integration.

**Methods**

**Fabrication.** The micro-disk is fabricated by photolithography assisted chemo-mechanical etching (PLACE) [38]. The fabrication process flow begins from the preparation of Erbium ion doped LN thin film by ion slicing. And the sample endures chromium (Cr) layer coating, hard mask patterning via femtosecond laser ablation, pattern transferring from the Cr hard mask to LN thin film via chemo-mechanical etching, and chemical wet etching to completely remove the Cr layer and partially remove the silica layer underneath the LN disk into the supporting pedestal. The details of the fabrication can be found in Ref. [17].

**Measurement.** To obtain the intensity distributions of up-conversion fluorescence, pump light and lasing signal, a set of short pass filters and long pass filters (Thorlabs, Inc., Model: FES800, FEL800, and FELH1100) were used to block the pump light, up-conversion fluorescence, and combination of the pump light and up-conversion fluorescence, respectively. To capture the intensity profiles of pump mode and up-conversion fluorescence signal in the micro-disk, a visible CCD is used to take the place of IR CCD in the optical micrograph system. The resolution of the OSA (YOKOGAWA, Inc., Model AQ6370D) was set as 0.01 nm.